\documentclass[a4paper,american]{scrartcl}
\usepackage{lmodern}
\usepackage[utf8]{inputenc}
\usepackage[T1]{fontenc}
\usepackage{color}
\usepackage{todonotes}
\usepackage{babel}
\usepackage{amsmath}
\usepackage{amssymb}
\usepackage{setspace}
\usepackage[authoryear]{natbib}
\usepackage[font=small]{quoting}
\usepackage{latexsym}
\usepackage[bookmarks=false,pdfborder={0 0 0},colorlinks=false]{hyperref}

\makeatletter

\newcounter{contatore}
\newcounter{contatore1}
\newcounter{contatore2}
\makeatother

\begin{document}

\title{Collapse or no collapse? What is the best ontology of quantum mechanics in the primitive ontology framework?}

\author{Michael Esfeld\thanks{University of Lausanne, Department of Philosophy,
    CH-1015 Lausanne, Switzerland. E-mail:
    \protect\href{mailto:michael-andreas.esfeld@unil.ch}{michael-andreas.esfeld@unil.ch}}
}
\maketitle
\begin{abstract}
      \begin{center} For Shan Gao (ed.), \emph{Collapse of the wave function},
      
      Cambridge University Press, forthcoming   \end{center}
    \medskip

Recalling the state of the art in the interpretation of quantum physics, this paper emphasizes that one cannot simply add a collapse parameter to the Schrödinger equation in order to solve the measurement problem. If one does so, one is also committed to a primitive ontology of a configuration of matter in physical space in order to have something in the ontology that constitutes the determinate measurement outcomes. The paper then argues that in the light of this consequence, the collapse postulate loses its attractiveness in comparison to an ontology of persisting particles moving on continuous trajectories according to a deterministic law.  
\medskip{}

\end{abstract}

\tableofcontents{}

\section{The state of the art: the measurement problem and beyond}

The measurement problem is the central issue in the formulation and understanding of any quantum theory, since it concerns the link between the theory and the data. The standard set up of this problem for non-relativistic quantum mechanics (QM) is provided by \citet[][p. 7]{Maudlin:1995aa}: \begin{quote}
(1) The wave-function of a system is \emph{complete}, i.e. the wave-function specifies (directly or indirectly) all of the physical properties of a system.

(2) The wave-function always evolves in accord with a linear dynamical equation (e.g. the Schrödinger equation).

(3) Measurements of, e.g., the spin of an electron always (or at least usually) have determinate outcomes, i.e., at the end of the measurement the measuring device is either in a state which indicates spin up (and not down) or spin down (and not up).
\end{quote}

Any two of these propositions are consistent with one another, but the conjunction of all three of them is inconsistent. This can be easily illustrated by means of Schrödinger's cat paradox (\cite{Schroedinger:1935aa}, p. 812): if the cat is completely described by the wave function and if the wave function always evolves according to the Schrödinger equation, then, due to the linearity of this wave equation, superpositions and entangled states will in general be preserved. Consequently, a measurement of the cat will in general not have a determinate outcome: at the end of the measurement, the cat will not be in the state of either being alive or being dead.

Hence, the measurement problem is not just a -- philosophical -- problem of the interpretation of a given formalism. It concerns also the very formulation of a consistent quantum theory. Even if one abandons (3), one has to put forward a formulation of quantum physics that establishes a link with at least the appearance of determinate measurement outcomes. If one retains (3), one has to develop a formulation of a quantum theory that goes beyond a theory in which only a wave function and a linear dynamical equation for the evolution of the wave function figure. Accordingly, the formulation of a consistent quantum theory can be divided into many worlds theories, rejecting (3), collapse theories, rejecting (2), and additional variable theories, rejecting (1).

However, research in the last decade has made clear that we do not face three equally distinct possibilities to solve the measurement problem, but just two: the main dividing line is between endorsing (3) and rejecting it. If one endorses (3), the consequence is not that one has to abandon \emph{either} (1) \emph{or} (2), but that one has to amend \emph{both} (1) \emph{and} (2). Determinate measurement outcomes as described in (3) are outcomes occurring in ordinary physical space, that is, in three-dimensional space or four-dimensional space-time. Hence, endorsing (3) entails being committed to the existence of a determinate configuration of matter in physical space that constitutes measurement outcomes (such as a live cat, or an apparatus configuration that indicates spin up, etc.). If one does so, one cannot stop at amending (2). The central issue then is not whether or not a collapse term for the wave function has to be added to the Schrödinger equation, because even with the addition of such a term, this equation still is an equation for the evolution of the wave function, by contrast to an equation for the evolution of a configuration of matter in physical space. Consequently, over and above the Schrödinger equation -- however amended -- a law or rule is called for that establishes an explicit link between the wave function and the configuration of matter in physical space. By the same token, (1) has to be changed in such a way that reference is made to the configuration of matter in physical space and not just the quantum state as encoded in the wave function.

Here and in the following, the configuration of matter in physical space is intended to be the \emph{entire} configuration of matter of the universe at a given time (recall that we are concerned with non-relativistic QM); accordingly, the wave function is intended to be the \emph{universal} wave function, that is, the wave function of the entire universe. In other words, the quantum theory we are after is a universal theory -- that is, a theory whose laws apply to the entire universe --, like classical mechanics. It is of course not the final theory, again like classical mechanics.  

In the literature, the configuration of matter in physical space is known as the \emph{primitive ontology} of quantum physics.\footnote{This term goes back to \citet[][ch. 2]{Durr:2013aa}, originally published 1992. A forerunner of this notion can be found in \citet[][p. 46]{Mundy:1989aa}. Cf. also Bell's notion of ``local beables'' in \citet[][ch. 7]{Bell:2004aa}, originally published 1975.} There are three proposals to spell out what that configuration is:

\begin{enumerate}
    \item \emph{Point particles} that consequently always have a determinate position in physical space (there cannot be point particles in physical space without these particles being located somewhere in that space, independently of whether or not we are able to determine that location). The corresponding quantum theory is \emph{Bohmian mechanics (BM)}, which adds particle positions to the quantum state as given by the wave function and a law for the evolution of these positions to the Schrödinger equation. That law is known as the guiding equation. The wave function figures in that law, its job being to determine the velocity of the particles at a time $t$, given their positions at $t$. This theory goes back to  \cite{Broglie:1928aa} and \cite{Bohm:1952aa}. Its dominant contemporary version is the one of \cite{Durr:2013aa} (and see the textbook \cite{Durr:2009fk}). 
    
    \item \emph{A matter density field} that stretches all over physical space, having varying degrees of density at different points or regions of space. The corresponding quantum theory is the  \emph{GRW matter density theory (GRWm)}, which uses the collapse formalism of Ghirardi, Rimini and Weber (\cite{Ghirardi:1986aa}) in the version of a continuous spontaneous localization (CSL) of the wave function in configuration space (\cite{Ghirardi:1990aa}). It establishes a link between the wave function in configuration space and the matter in physical space in such a way that the wave function and its evolution according to the GRW-CSL equation describe a wave or field filling all of physical space and the evolution of that field in physical space (\cite{Ghirardi:1995aa}). 
    
    \item \emph{Isolated point events in physical space, called ``flashes''}. These flashes are ephemeral. They do not stretch beyond the space-time point at which they occur. The corresponding quantum theory is the \emph{GRW flash theory (GRWf)}, which uses the original collapse formalism of Ghirardi, Rimini and Weber (\cite{Ghirardi:1986aa}) with the wave function making occasionally jumps so that it undergoes a spontaneous localization in configuration space. A rule then is added to the GRW equation stating that whenever a spontaneous localization of the wave function occurs in configuration space, a flash shows up at a point in physical space. The flash theory has been proposed as an ontology of the GRW formalism by \citet[][ch. 22, originally published 1987]{Bell:2004aa}. The term ``flash'' was coined by \citet[][p. 826]{Tumulka:2006aa}.  
\end{enumerate}         

Despite their differences, these three quantum theories have a number of important features in common (\cite{Allori:2008aa} were the first to work out their common structure). In the first place, (a) they are all committed to the matter in physical space being primitive objects. That is to say, these objects do not have any physical properties over and above their being localized in physical space. Hence, by contrast to what one can maintain about classical particles, these objects cannot be considered as having an intrinsic mass or an intrinsic charge (or an intrinsic spin)\footnote{See \citet[][ch. 4]{Bell:2004aa}, originally published 1971, and \cite{Norsen:2014aa} for the Bohmian treatment of spin. Similar remarks apply to the GRWm and GRWf theories.}. They do not have any intrinsic properties. As regards BM, experimental considerations involving interference phenomena -- for instance in the context of the Aharonov-Bohm effect and of certain interferometry experiments -- show, in brief, that mass and charge are effective at all the possible particle locations that the wave function admits; they can hence not be considered as properties that are localized at the Bohmian particle positions (see e.g. \cite{Brown:1995aa, Brown:1996aa} and references therein; cf. also most recently \cite{Pylkkanen:2014aa}). In brief, any property that one may contemplate attributing to the particles over and above their position is in fact situated at the level of their quantum state as represented by the wave function instead of being a candidate for an intrinsic property of the particles. The same goes for the flash-events on the GRWf theory. When it comes to the GRWm theory, the matter density field is a primitive stuff filling all of space, which admits different degrees of density as a primitive matter of fact, but which has no further physical properties, as \cite{Allori:2013aa} point out: \begin{quote}
Moreover, the matter that we postulate in GRWm and whose density is given by the $m$ function does not ipso facto have any such properties as mass or charge; it can only assume various levels of density. (\cite{Allori:2013aa}, pp. 331--332)
\end{quote}

Furthermore, (b) the fact that all the primitive ontology theories of quantum physics infringe upon proposition (1) of the measurement problem comes out clearly when one enquires into their consequences for our knowledge: as \cite{Cowan:2015aa} establish, not only the Bohmian particle positions -- and hence the particle trajectories -- are not always accessible to an observer, but also the GRWm matter density field and the GRWf flash distribution are not entirely accessible to an observer, although the latter are specified by the wave function, whereas the Bohmian particle positions are not specified by the wave function. Hence, one does not avoid what may seem to be a drawback of subscribing to so-called hidden variables, namely a limited epistemic accessibility, by amending the Schrödinger equation and taking the wave function as it figures in such an amended Schrödinger equation to represent the configuration of matter in physical space. In a nutshell, if one is committed to a configuration of matter in physical space in quantum physics, one also has to endorse a commitment to a limited epistemic accessibility of that configuration, in whatever way one spells out the theory of that configuration and its evolution.

By way of consequence, (c) probabilities enter into any primitive ontology theory of QM through our ignorance of the exact configuration of matter in physical space. This ignorance implies that we immediately have to resort to probabilistic descriptions in QM, independently of whether the dynamical law for the evolution of the configuration of matter in physical space is deterministic (as in Bohmian mechanics) or stochastic (as in the GRW theory) (see e.g. \cite{Oldofredi:2016aa} on that introduction of probabilities).

Finally, (d) endorsing a primitive ontology theory of quantum physics does not commit us to subscribing to an ontological dualism of matter in physical space (the primitive ontology, consisting in primitive objects) and the quantum state (the wave function) in configuration space. Such a dualism would be highly implausible; for instance, it would be unintelligible how a wave function, being a field on configuration space, could interact with matter in physical space by influencing its motion. There are several proposals in the literature how to conceive the wave function in the primitive ontology framework that all avoid such an implausible dualism: (i) The most straightforward proposal is to ban the wave function as an additional entity from the ontology altogether. The claim thus is that the primitive ontology is the entire ontology. Given the spatial distribution of the elements of the primitive ontology throughout the whole history of the universe, the universal wave function and its evolution is part of the best descriptive system, that is, the system that achieves the best balance between being simple and being informative about that distribution. The wave function hence has only a descriptive role by contrast to an ontological one. This stance is known as quantum Humeanism. It has been worked out in the philosophy literature mainly by taking BM as an example (\cite{Miller:2014aa}, \cite{Esfeld:2014aa}, \cite{Callender:2014aa}, \cite{Bhogal:2015aa}), but there also exists a concrete model in the physics literature that uses the GRWf theory (\cite{Dowker:2005aa}). (ii) When one admits the wave function to the ontology, one can still regard it as referring to an entity in physical space. The most prominent proposal in that respect in the philosophical literature is to take the wave function to refer to dispositions of the elements of the primitive ontology for a certain evolution under certain circumstances, namely either one holistic disposition of the entire configuration (\cite{Esfeld:2014ab}), or a multitude of dispositions of each individual element whose manifestations depend on the other elements (\cite{Suarez:2015aa}). (iii) The corresponding proposal in the physics literature, conceiving the wave function as an entity in physical space, is the one of regarding it either as a multi-field in physical space -- that is, a field that attributes a value not to single space-time points, but only to an entire bunch of them (\cite{Forrest:1988aa}, ch. 6.2) -- or as referring to a multitude of fields associated with each element of the primitive ontology (such as each Bohmian particle) (\cite{Norsen:2015aa}).  

Given this state of the art, the next task then is to evaluate the theories that solve the measurement problem by being committed to a primitive ontology of matter distributed in physical space. That is, one seeks for an answer to the following two questions: What is best proposal for the physical objects? What is best proposal for the dynamics? This paper sets out to answer these two questions (see also \cite{Esfeld:2014ac}).

\section{What is best proposal for the physical objects?}

Atomism is the oldest and most influential tradition in natural philosophy, going back to the pre-Socratic philosophers Leucippus and Democritus and having been turned into a precise physical theory by Newton. Atomism offers a clear and simple explanation of the realm of our experience. Macroscopic objects are composed of point particles. All the differences between the macroscopic objects -- at a time as well as in time -- are accounted for in terms of the spatial configuration of these particles and its change, which is subject to certain laws. That is why Feynman famously writes at the beginning of the \emph{Feynman lectures on physics}: \begin{quote} If, in some cataclysm, all of scientific knowledge were to be destroyed, and only one sentence passed on to the next generations of creatures, what statement would contain the most information in the fewest words? I believe it is the \emph{atomic hypothesis} (or the atomic \emph{fact}, or whatever you wish to call it) that \emph{all things are made of atoms -- little particles that move around in perpetual motion, attracting each other when they are a little distance apart, but repelling upon being squeezed into one another.} In that one sentence, you will see, there is an enormous amount of information about the world, if just a little imagination and thinking are applied. (\cite{Feynman:1963aa}, ch. 1-2)\end{quote}

When it comes to quantum physics, atomism -- that is, an ontology of permanent particles moving on continuous trajectories -- is implemented in BM. What changes in BM with respect to classical mechanics is the law of motion for these particles, namely a non-local law that takes quantum entanglement into account.

There are three main reasons to take an ontology of particles to be the best proposal also in the domain of quantum physics: (i) In the first place, also in this domain, all experimental evidence is evidence of discrete objects (i.e. particles) -- from dots on a display to traces in a cloud chamber. Entities that are not particles -- such as waves or fields -- come in as figuring in the explanation of the behaviour of the particles, but they are not themselves part of the experimental evidence. For instance, the double slit experiment is made apparent by particles hitting on a screen. (ii) The argument from composition is not touched by the transition from classical to quantum physics: from the chemical elements on to all macroscopic objects, everything is composed of elementary quantum particles. (iii) Consequently, all other proposals are parasitic on the particle proposal: even if they do not admit particles in the ontology -- such as the GRWm theory --, their formalism for QM works in terms of a fixed number of $N$ permanent particles, which defines the configuration space; its dimension is $3N$, with each point of that space corresponding to a possible particle configuration in physical space. Hence, these other proposals have to interpret a particle configuration formalism as not representing a particle configuration.

However, why should one retain particles and thus trajectories in the light of the evidence from quantum physics? The mentioned arguments speak in favour of an ontology of discrete objects, but one may wonder whether this has to be an ontology of particles. The GRWf ontology of single, discrete events can be considered as a particle ontology without the trajectories so that what remains of the particles are isolated events in space-time. However, the three above mentioned arguments can be adapted in such a way that they single out particles over flashes: (iii) applies not only to the GRWm ontology, but also to the GRWf ontology: the flashes are discontinuous, and there is no intertemporal identity of them, since the flashes do not persist in time. But the formalism is based on a fixed number of persistent particles. (ii) This observation applies also to the argument from composition: there is nothing in the GRWf ontology that could ground the intertemporal identity of macroscopic objects.

As regards (i), it is true that the particle evidence in quantum physics is evidence of discrete objects, which could be point events like point particles. But the evidence of discrete objects in physical space is not evidence of a physical space into which discrete objects are inserted; it is evidence only of relative positions of discrete objects, that is, distances among discrete objects. However, by renouncing on persisting objects, the GRWf ontology is committed to the existence of an absolute background space into which the flashes are inserted and an absolute background time in which the flashes show up (there can even be times at which there are no flashes at all in the universe). By contrast, although BM is usually also formulated in terms of the particles moving in a background space and a background time, the ontological commitment to a background space and time can be dispensed with in BM. The ontology of BM can be conceived only in terms of distance relations among particles that change, with the representation of that change in terms of particle trajectories in a background space and time being only a means of representation of the primitive ontology instead of implying the ontological commitment to an absolute space and time (see \cite{Vassallo:2016ac} and \cite{Vassallo:2016aa}). By the same token, one can regard the universal wave function in BM as a means of representation only instead of as an element of the ontology over and above the particles, as mentioned at the end of the preceding section.       

The ontology of matter being one continuous stuff, known as gunk, that fills all of space is as old as the ontology of atomism, going back to the Presocratic natural philosophers as well. This view does not have to commit itself to points, neither to material points nor to points of space (see \cite{Arntzenius:2005aa}). It is hence not tied to endorsing an absolute background space: it can be construed as being committed to a continuous stuff that is extended, but not to an absolute space that is distinct from that stuff and into which that stuff is inserted.  

In order to accommodate variation, gunk cannot be conceived as being homogeneous throughout space. To take variation into account, one has to maintain that there is more stuff in some parts of space and less stuff in others. Atomism conceptualizes variation in terms of different distances among the discrete point particles so that some particles are situated close to one another, whereas others are further apart: there are clusters of point particles with distances among them that are smaller than the distances that these particles bear to particles outside such a cluster. By contrast, the gunk ontology cannot accommodate variation throughout space in terms of the concentration of primitive point particles. It therefore faces this question: What constitutes the fact of there being more matter in some regions of space and less matter in others?

The view of matter being gunk has to acknowledge as a further primitive a variation of the density of gunk throughout space with gunk being more dense in some parts of space and less dense in other parts. That is to say, gunk admits of degrees, as expressed by the $m$ function in the GRWm formalism: there is more stuff in some parts of space than in others, with the density of matter in the parts of space changing in time; otherwise, the theory would not be able to accommodate variation. Formally, one can represent the degrees of density in terms of attributing a value of matter density to the points of space (the $m$ function as evaluated at the points of space), although the matter density stuff, being gunk, is infinitely divisible, and this ontology is not committed to the existence of points of space. The main problem is that it remains unclear what could constitute the difference in degrees of stuff at points of space, if matter just is primitive stuff. The gunk theory thus is committed to the view of matter being a bare substratum with its being a primitive fact that this substratum has various degrees of density in different parts of space. In a nutshell, there is a primitive stuff-essence of matter that furthermore admits of different degrees of density. In comparison to the gunk view of matter, atomism is the simpler and clearer proposal for an ontology of fundamental physics, because it avoids the dubious commitment to a bare substratum or primitive stuff-essence of matter with different degrees of density.

It seems that QM favours a particle ontology, given that the formalism is conceived in terms of a fixed number of permanent particles, but that quantum field theory (QFT) clearly favours the view of matter being a field and thus one continuous stuff filling all of space instead of discrete objects. However, a quantum field never is a field in the sense of a continuous entity filling space that has definite values at the points of space. The fields figuring in the textbook formalism of QFT are operator valued fields, that is, mathematical objects employed to calculate probabilities for obtaining certain measurement outcomes at certain points of space if certain procedures are applied. They hence do not represent properties that occur at points of space or space-time. No ontology can be built on operator valued fields. As regards the experimental evidence, the remarks above apply, namely that also in the domain of QFT, all the experimental evidence is one of discrete objects, such as particle traces in a cloud chamber. When it comes to the wave function, again, like in QM, it is a field on configuration space and not a continuous stuff in physical space.

Consequently, the measurement problem hits QFT in the same way as QM (see \cite{Barrett:2014aa}). In particular, there is no relativistic theory of measurement. Consequently, problems that primitive ontology theories of quantum physics may face with respect to relativistic physics cannot be counted as an argument against these theories: they solve the measurement problem, and no one has produced a solution to this problem that (a) acknowledges determinate measurement outcomes and (b) is a relativistic theory of interactions, including in particular measurement interactions (see again \cite{Barrett:2014aa}). BM has to rely on a privileged foliation of space-time into spatial hyersurfaces in its ontology, which can be introduced through the universal wave function (see \cite{Duerr:2013}). However, this foliation does not show up in the deduction of the statistics of measurement outcomes from the dynamical laws. As regards GRWf and GRWm, there are relativistic versions as long as one considers only a distribution of the elements of the primitive ontology throughout the whole of space-time (see \cite{Tumulka:2006aa} for GRWf and \cite{Bedingham:2014aa} for GRWm), but no relativistic versions of interactions (see \cite{Esfeld:2014ad}).   

The Bohmian solution of the quantum measurement problem works for QFT in the same way as for QM: as it is a \emph{non sequitur} to take particle trajectories to be ruled out in QM due to the Heisenberg uncertainty relations, so it is a \emph{non sequitur} to take permanent particles moving on definite trajectories according to a deterministic law to be ruled out in QFT due to the statistics of particle creation and annihilation phenomena. In both cases, such an underlying particle ontology is in the position to explain the statistics of measurement outcomes (see \cite{Colin:2007aa} and \cite{Deckert:2016ac} as to how this is achieved in a Bohmian QFT in the framework of what is known as the Dirac sea model). In brief, QFT does not change anything with respect to the evaluation of the proposals for a primitive ontology of quantum physics: the arguments that favour a particle ontology remain valid in the domain of QFT.

Nonetheless, atoms \emph{qua} point particles are theoretical entities. They are not seen by the naked eye when one sees, for instance, dots on a screen as outcomes of the double slit experiment. They are admitted because they provide the best explanation of the observable facts. The simplicity and parsimony of this proposal are part of the case for its being the best explanation. To put it in a nutshell, \emph{particle evidence is best explained in terms of particle ontology}. However, this explanation is not given by the ontology of point particles alone, but by this ontology together with the dynamics that is put forward to describe the motion of the particles: it is the dynamics that provides for the stability of the macroscopic objects with which we are familiar. That is why assessing the proposals for a primitive ontology of quantum physics depends not only on the ontology put forward for the physical objects, but also on the dynamics that is conceived for these objects.

\section{What is best proposal for the dynamics?}

BM is usually presented by formulating the laws of motion on the configuration space $\mathbb{R}^{3N}$, where $N$ is the number of particles and $x_1,\dots,x_k,\dots,x_N \in \mathbb{R}^{3N}$ represents their positions at time $t$. The configuration then evolves according to the guiding equation 
\begin{equation}\label{eq:standard-guiding-equation}
\frac{\mathrm{d}x_k}{\mathrm{d}t}= \mathrm{Im}  \frac{\hbar}{m_k}\frac{\psi^*\nabla_k \psi}{\psi^* \psi} (x_1,\dots , x_N),
\end{equation}
where $\psi(x_1,\dots, x_n)$ is the wave-function representing the quantum state of the system and $\mathrm{Im}$ denotes the imaginary part. The time-evolution of this wave-function, in turn, is given by the  Schrödinger equation
\begin{equation}
\imath\hbar\frac{\partial\psi}{\partial t}= \Bigl(-\sum_{j=1}^N \frac{\hbar^2}{2m_j} \Delta_j + V(x_1,\dots , x_n) \Bigr)\, \psi,\label{eq:Schroedinger-equation}
\end{equation}
familiar from standard QM. The non-local character of the guiding equation is manifested in the fact that the velocity of any particle at time $t$ depends on the position of every other particle at time $t$;  the law of motion, in other words, describes the evolution of the particle configuration \textit{as a whole}. This is necessary in order to take quantum non-locality -- as illustrated for instance by Bell's theorem -- into account.

In GRW, the evolution of the wave function $\Psi_t$ is given by a modified Schrödinger equation. The latter can be defined as follows: the wave function undergoes spontaneous jumps in configuration space at random times distributed according to the Poisson distribution with rate $N\lambda$. Between two
successive jumps the wave function $\Psi_t$ evolves according to the usual Schrödinger equation. At the time of a jump the $k$th component of the wave function $\Psi_t$ undergoes an instantaneous collapse according to
\begin{align}
    \Psi_t(x_1,\dots,x_k,\dots,x_N) \mapsto \frac{(L^x_{x_k})^{1/2}
\Psi_t(x_1,\dots,x_k,\dots,x_N)}{\|(L^x_{x_k})^{1/2} \Psi_t\|}, 
\end{align}
where the localization operator $L^x_{x_k}$ is given as a multiplication operator of the form
\begin{align}
    L^x_{x_k} := \frac{1}{(2\pi\sigma^2)^{3/2}} e^{-\frac{1}{2\sigma^2}(x_k-x)^2},
\end{align}

and $x$, the centre of the collapse, is a random position distributed according to the probability density $p(x)=\|(L^x_{x_k})^{1/2} \Psi_t\|^2$. This modified Schrödinger evolution captures in a mathematically precise way what the collapse postulate in textbook QM introduces by a \emph{fiat}, namely the collapse of the wave function so that it can represent localized objects in physical space, including in particular measurement outcomes. GRW thereby introduce two additional parameters, the mean rate $\lambda$ as well as the width $\sigma$ of the localization operator, which can be regarded as new constants of nature whose values can be inferred from (or are at least bounded by) experiments (such as chemical reactions on a photo plate, double slit experiments, etc.). An accepted value of the mean rate $\lambda$ is of the order of $10^{15}s^{-1}$. This value implies that the spontaneous localization process for a single particle occurs only at astronomical time scales of the order of $10^{15}s$, while for a macroscopic system of $N\sim10^{23}$ particles, the collapse happens so fast that possible superpositions are resolved long before they would be experimentally observable. Moreover, the value of $\sigma$ can be regarded as localization width; an accepted value is of the order of $10^{-7}m$. The latter is constrained by the overall energy increase of the wave function of the universe that is induced by the localization processes.

However, as explained in the first section, modifying the Schrödinger equation is, by itself, not sufficient to solve the measurement problem: to do so, one has to answer the question of what the wave function and its evolution represent. One therefore has to add to the GRW equation a link between the evolution of the mathematical object $\Psi_t$ in configuration space and the distribution of matter in physical space in order to account for the outcomes of experiments and, in general, the observable phenomena. \cite{Ghirardi:1995aa} accomplish this task by taking the evolution of the wave function in configuration space to represent the evolution of a matter density field in physical space. This then constitutes the GRWm theory. It amounts to introduce in addition to $\Psi_t$ and its time evolution a field $m_t(x)$ on physical space $\mathbb R^3$ as follows:
\begin{align}
    m_t(x) = \sum_{k=1}^N m_k \int d^3x_1\dots d^3x_N\, \delta^3(x-x_k)
    |\Psi_t(x_1,\dots,x_N)|^2.
\end{align}
This field $m_t(x)$ is to be understood as the density of matter in physical space $\mathbb R^3$ at time $t$ (see \cite{Allori:2008aa}, section 3.1).  

By introducing two new dynamical parameters -- lambda and sigma -- whose values have to be put in by hand, the GRW theory abandons the simplicity and elegance of the Schrödinger equation and the Bohmian guiding equation, without amounting to a physical benefit (there is of course a benefit in comparison to stipulating the collapse postulate by a simple \emph{fiat}, but doing so is no serious theory). Indeed, there is an ongoing controversy whether the GRWm ontology of a continuous matter density field that develops according to the GRW equation is sufficient to solve the measurement problem. The reason is the so-called problem of the tails of the wave function. This problem arises from the fact that the GRW theory mathematically implements spontaneous localization by multiplying the wave function with a Gaussian, such that the collapsed wave function, although being sharply peaked in a small region of configuration space, does not actually vanish outside that region; it has tails spreading to infinity. In the literature starting with \cite{Albertandloewer:1996aa} and P. \cite{Lewis:1997aa}, it is therefore objected that the GRW theory does not achieve its aim, namely to describe measurement outcomes in the form of macrophysical objects having a determinate position. However, there is nothing indefinite about the positions of objects according to GRWm. It is just that an (extremely small) part of each object’s matter is spread out through all of space. But since the overwhelming part of any ordinary object’s matter is confined to a reasonably small spatial region, we can perfectly well express this in our (inevitably vague) everyday language by saying that the object is in fact located in that region (see \cite{Monton:2004aa}, pp. 418-419, and \cite{Tumulka:2011aa}). Thus, the GRWm ontology offers a straightforward solution to what \citet[][p. 56]{Wallace:2008aa} calls the problem of bare tails.

However, there is another aspect, which is known as the problem of structured tails (see \cite{Wallace:2008aa}, p. 56). Consider a situation in which the pure Schrödinger evolution would lead to a superposition with equal weight of two macroscopically distinct states (such as a live and a dead cat). The GRW dynamics ensures that the two weights do not stay equal, but that one of them (e.g. the one pertaining to the dead cat) approaches unity while the other one becomes extremely small (but not zero). In terms of matter density, we then have a high-density dead cat and a low-density live cat. The problem is that it seems that the low-density cat is just as cat-like (in terms of shape, behaviour, etc.) as the high-density cat, so that in fact there are two cat-shapes in the matter density field, one with a high and another one with a low density. There is an ongoing controversy about this problem: \citet[][pp. 135-138]{Maudlin:2010aa} takes it to be a knock down objection against the GRW matter density ontology, whereas others put forward reasons that aim at justifying to dismiss the commitment to there being a low-density that is as cat-like as the high-density cat in the matter density field (see notably \cite{Wallace:2014aa}, \cite{Albert:2015aa}, pp. 150-154, and \cite{Egg:2015aa}, section 3).

Be that as it may, there arguably is another, more important drawback of the GRW dynamics that concerns the meaning of the spontaneous localization of the wave function in configuration space for the evolution of the matter density field in physical space. To illustrate this issue, consider a simple example, namely the thought experiment of one particle in a box that Einstein presented at the Solvay conference in 1927 (the following presentation is based on de Broglie’s version of the thought experiment in \cite{Broglie:1964aa}, pp. 28-29, and on \cite{Norsen:2005aa}): the box is split in two halves which are sent in opposite directions, say from Brussels to Paris and Tokyo. When the half-box arriving in Tokyo is opened and found to be empty, there is on all accounts of QM that acknowledge that measurements have outcomes a fact that the particle is in the half-box in Paris.

On GRWm, the particle is a matter density field that stretches over the whole box and that is split in two halves of equal density when the box is split, these matter densities travelling in opposite directions. Upon interaction with a measurement device, one of these matter densities (the one in Tokyo in the example given above) vanishes, while the matter density in the other half-box (the one in Paris) increases so that the whole matter is concentrated in one of the half-boxes. One might be tempted to say that some matter travels from Tokyo to Paris; however, since it is impossible to assign any finite velocity to this travel, the use of the term ``travel'' is inappropriate. For lack of a better expression let us say that some matter is delocated from Tokyo to Paris (this term has been proposed by Matthias Egg, see \cite{Egg:2014aa}, p. 193); for even if the spontaneous localization of the wave function in configuration space is conceived as a continuous process as in \cite{Ghirardi:1990aa}, the time it takes for the matter density to disappear in one place and to reappear in another place does not depend on the distance between the two places. This delocation of matter, which is not a travel with any finite velocity, is quite a mysterious process that the GRWm ontology asks us to countenance.

On BM, by contrast, in this example, there always is one particle moving on a continuous trajectory in one of the two half-boxes, and opening one of them only reveals where the particle was all the time. In other words, BM provides a local account of the case of the particle in a box. However, when moving from Einstein’s thought experiment with one particle in a box (1927) to the EPR experiment (\cite{Einstein:1935aa}), even BM can no longer give a local account, as proven by Bell's theorem (\cite{Bell:2004aa}, ch. 2; see also notably chs. 7 and 24). On the GRWm theory, again, the measurement in one wing of the experiment triggers a delocation of the matter density, more precisely a change in its shape in both wings of the experiment, so that, in the version of the experiment by \citet[][pp. 611-622]{Bohm:1951aa} the shape of the matter density constitutes two spin measurement outcomes. On BM, fixing the parameter in one wing of the EPR experiment influences the trajectory of the particles in both wings via the wave function of the whole system, which consists of the measured particles as well as of the particles that make up the measuring devices.

Hence, in this case, it clearly comes out that according to the Bohmian velocity equation \eqref{eq:standard-guiding-equation}, the velocity of any particle depends strictly speaking on the position of all the other particles. However, each particle always moves with a determinate, finite velocity that is not greater than the velocity of light so that its motion traces out a continuous trajectory, without anything jumping -- or being delocated -- in physical space. The best conjecture for a velocity field that captures this motion that we can make, namely \eqref{eq:standard-guiding-equation}, requires acknowledging that the motions of these particles are correlated with each other, but this does not imply a commitment to there being some spooky agent or force in nature that instantaneously coordinates the motions of all the particles in the universe. Quantum physics just teaches us that it is a fact about the universe that when we seek to write down a simple and general law that accounts for the empirical evidence, we have to conceive a law that represents the motions of the particles to be correlated with one another.

However, \cite{Einstein:1948aa} is certainly right in pointing out that a complete suspension of the principles of separability and local action would make it impossible to do physics: a theory that says that the motion of any object is effectively influenced by the position of every other object in the configuration of matter of the universe would be empirically inadequate and rule out any experimental investigation of nature. In order to meet Einstein's requirement, it is not necessary to rely on a collapse dynamics, as does GRW. BM fulfills this condition because decoherence will in general destroy the entanglement between large and/or distant systems, allowing to treat them, for all practical purposes, as evolving in an independent manner. Moreover, BM is able to recover classical behaviour in the relevant regimes (see \cite{Durr:2013aa}, ch. 5). Since BM is a theory about the motion of particles, this classical limit does not involve or require any change in the ontological commitment, but consists in the proposition that typical Bohmian trajectories look approximately Newtonian on macroscopic scales (if the characteristic wave length associated to $\psi$ is small compared to the scale on which the interaction potential varies). Altogether, the Bohmian theory, against the background of an ontology of point particles that are characterized only by their relative positions and a dynamics for the change of these positions, illustrates that there is nothing suspicious about a non-local dynamics.

Let us turn now to GRWf. As mentioned in the preceding section, the flashes can be conceived as Bohmian particles deprived of their trajectories, so that all that is left are isolated point events in space-time. The main problem for this ontology is that the flashes are too sparsely distributed. Consider what this means for the dynamics: the account that the original GRW theory envisages for measurement interactions does not work on the flash ontology -- in other words, this ontology covers only the spontaneous appearance and disappearance of flashes, but offers no account of interactions. On the original GRW proposal, a measurement apparatus is supposed to interact with a quantum object; since the apparatus consists of a great number of quantum objects, the entanglement of the wave function between the apparatus and the measured quantum object will be immediately reduced due to the spontaneous localization of the wave function of the apparatus. However, even if one supposes that a measurement apparatus can be conceived as a galaxy of flashes (but see the reservations of \cite{Maudlin:2011aa}, pp. 257-258), there is on GRWf nothing with which the apparatus could interact: there is no particle that enters it, no mass density and in general no field that gets in touch with it either (even if one conceives the wave function as a field, it is a field in configuration space and not a field in physical space where the flashes are). There only is one flash (standing for what is usually supposed to be a quantum particle) in its past light cone, but there is nothing left of that flash with which the apparatus could interact.

\section{Conclusion}

This paper has made evident that one cannot simply add a collapse parameter to the Schrödinger equation in order to solve the measurement problem. If one does so, one also has to specify a primitive ontology of a configuration of matter in physical space in order to have something in the ontology that constitutes the determinate measurement outcomes and that evolves as described by the modified Schrödinger equation. This then is the important ingredient for the ontology and not the wave function, independently of whether or not it undergoes collapses. As explained at the end of section 1, one can endorse the wave function as a parameter that figures centrally in the law of motion for the primitive ontology, but nevertheless regard it only as a bookkeeping device of that motion.

In the light of this consequence, the collapse principle loses its attractiveness. If one needs a primitive ontology over and above the collapse postulate anyway, one can retain a particle ontology and a deterministic law of motion for the particles with a universal wave function that never collapses and deduce the QM probability calculus from that law. Recall that, as mentioned in section 1, any primitive ontology implies that we do not enjoy full epistemic accessibility to the configuration of matter, so that probabilities come in anyway through our ignorance of the exact initial conditions, independently of whether or not the law for the evolution of the wave function is stochastic. Of course, this assessment would change if experimental tests of collapse theories like GRW against theories that exactly produce the predictions of textbook QM -- such as BM -- were carried out successfully and confirmed the collapse theories where they deviate from the standard predictions (see \cite{Curceanu:2016aa} for such experiments). 

As things stand, the arguments for the particle ontology notably are that all the experimental evidence is particle evidence, that all composed objects are made of particles and that any QM formalism is conceived in terms of a definite number of persisting particles. As regards the dynamics, a law for the particle motion on continuous trajectories such as the Bohmian guiding equation gives an account of the non-local correlations as brought out by Bell's theorem and the EPR experiment in terms of correlated particle motion without anything ever jumping or being delocated in physical space. 

\bibliographystyle{apalike}
\bibliography{references_fundont}

\begin{thebibliography}{}

\bibitem[Albert, 2015]{Albert:2015aa}
Albert, D.~Z. (2015).
\newblock {\em After physics}.
\newblock Cambridge, Massachusetts: Harvard University Press.

\bibitem[Albert and Loewer, 1996]{Albertandloewer:1996aa}
Albert, D.~Z. and Loewer, B. (1996).
\newblock Tails of {S}chr{\"o}dinger's cat.
\newblock In Clifton, R.~K., editor, {\em Perspectives on quantum reality},
  pages 81--91. Dordrecht: Kluwer.

\bibitem[Allori et~al., 2008]{Allori:2008aa}
Allori, V., Goldstein, S., Tumulka, R., and Zangh{\`\i}, N. (2008).
\newblock On the common structure of {B}ohmian mechanics and the
  {G}hirardi-{R}imini-{W}eber theory.
\newblock {\em British Journal for the Philosophy of Science}, 59(3):353--389.

\bibitem[Allori et~al., 2014]{Allori:2013aa}
Allori, V., Goldstein, S., Tumulka, R., and Zangh{\`\i}, N. (2014).
\newblock Predictions and primitive ontology in quantum foundations: a study of
  examples.
\newblock {\em British Journal for the Philosophy of Science}, 65(2):323--352.

\bibitem[Arntzenius and Hawthorne, 2005]{Arntzenius:2005aa}
Arntzenius, F. and Hawthorne, J. (2005).
\newblock Gunk and continuous variation.
\newblock {\em The Monist}, 88:441--465.

\bibitem[Barrett, 2014]{Barrett:2014aa}
Barrett, J.~A. (2014).
\newblock Entanglement and disentanglement in relativistic quantum mechanics.
\newblock {\em Studies in History and Philosophy of Modern Physics},
  48:168--174.

\bibitem[Bedingham et~al., 2014]{Bedingham:2014aa}
Bedingham, D., D\"{u}rr, D., Ghirardi, G.~C., Goldstein, S., Tumulka, R., and
  Zangh{\`\i}, N. (2014).
\newblock Matter density and relativistic models of wave function collapse.
\newblock {\em Journal of Statistical Physics}, 154:623--631.

\bibitem[Bell, 2004]{Bell:2004aa}
Bell, J.~S. (2004).
\newblock {\em Speakable and unspeakable in quantum mechanics}.
\newblock Cambridge: Cambridge University Press, second edition.

\bibitem[Bhogal and Perry, 2016]{Bhogal:2015aa}
Bhogal, H. and Perry, Z.~R. (2016).
\newblock What the {H}umean should say about entanglement.
\newblock {\em No{\^u}s}, page DOI 10.1111/nous.12095.

\bibitem[Bohm, 1951]{Bohm:1951aa}
Bohm, D. (1951).
\newblock {\em Quantum theory}.
\newblock Englewood Cliffs: Prentice-Hall.

\bibitem[Bohm, 1952]{Bohm:1952aa}
Bohm, D. (1952).
\newblock A suggested interpretation of the quantum theory in terms of
  ``hidden'' variables. 1.
\newblock {\em Physical Review}, 85(2):166--179.

\bibitem[Brown et~al., 1995]{Brown:1995aa}
Brown, H.~R., Dewdney, C., and Horton, G. (1995).
\newblock Bohm particles and their detection in the light of neutron
  interferometry.
\newblock {\em Foundations of Physics}, 25(2):329--347.

\bibitem[Brown et~al., 1996]{Brown:1996aa}
Brown, H.~R., Elby, A., and Weingard, R. (1996).
\newblock Cause and effect in the pilot-wave interpretation of quantum
  mechanics.
\newblock In Cushing, J.~T., Fine, A., and Goldstein, S., editors, {\em Bohmian
  mechanics and quantum theory: an appraisal}, volume 184 of {\em Boston
  Studies in the Philosophy of Science}, pages 309--319. Dordrecht: Springer.

\bibitem[Callender, 2015]{Callender:2014aa}
Callender, C. (2015).
\newblock One world, one beable.
\newblock {\em Synthese}, 192(10):3153--3177.

\bibitem[Colin and Struyve, 2007]{Colin:2007aa}
Colin, S. and Struyve, W. (2007).
\newblock A dirac sea pilot-wave model for quantum field theory.
\newblock {\em Journal of Physics A}, 40(26):7309--7341.

\bibitem[Cowan and Tumulka, 2016]{Cowan:2015aa}
Cowan, C.~W. and Tumulka, R. (2016).
\newblock Epistemology of wave function collapse in quantum physics.
\newblock {\em British Journal for the Philosophy of Science}, 67:405--434.

\bibitem[Curceanu and alteri, 2016]{Curceanu:2016aa}
Curceanu, C. and alteri (2016).
\newblock Spontaneously emitted x-rays: an experimental signature of the
  dynamical reduction models.
\newblock {\em Foundations of Physics}, 46(3):263--268.

\bibitem[de~Broglie, 1928]{Broglie:1928aa}
de~Broglie, L. (1928).
\newblock La nouvelle dynamique des quanta.
\newblock {\em Electrons et photons. Rapports et discussions du cinqui{\`e}me
  Conseil de physique tenu {\`a} Bruxelles du 24 au 29 octobre 1927 sous les
  auspices de l'Institut international de physique Solvay}, pages 105--132.
\newblock Paris: Gauthier-Villars. English translation in Bacciagaluppi, G. and
  Valentini, A., editors (2009). \emph{Quantum theory at the crossroads.
  Reconsidering the 1927 Solvay conference}, pages 341--371. Cambridge:
  Cambridge University Press.

\bibitem[de~Broglie, 1964]{Broglie:1964aa}
de~Broglie, L. (1964).
\newblock {\em The current interpretation of wave mechanics. A critical study}.
\newblock Amsterdam: Elsevier.

\bibitem[Deckert et~al., 2016]{Deckert:2016ac}
Deckert, D.-A., Esfeld, M., and Oldofredi, A. (2016).
\newblock A persistent particle ontology for {QFT} in terms of the {D}irac sea.
\newblock {\em British Journal for the Philosophy of Science}, pages Preprint
  http://philsci--archive.pitt.edu/12375/, arXiv:1608.06141[physics.hist--ph].

\bibitem[Dowker and Herbauts, 2005]{Dowker:2005aa}
Dowker, F. and Herbauts, I. (2005).
\newblock The status of the wave function in dynamical collapse models.
\newblock {\em Foundations of Physics Letters}, 18:499--518.

\bibitem[D{\"u}rr et~al., 2013a]{Duerr:2013}
D{\"u}rr, D., Goldstein, S., Norsen, T., Struyve, W., and Zangh{\`\i}, N.
  (2013a).
\newblock Can {B}ohmian mechanics be made relativistic?
\newblock {\em Proceedings of the Royal Society A}, 470:2162.

\bibitem[D{\"u}rr et~al., 2013b]{Durr:2013aa}
D{\"u}rr, D., Goldstein, S., and Zangh{\`\i}, N. (2013b).
\newblock {\em Quantum physics without quantum philosophy}.
\newblock Berlin: Springer.

\bibitem[D{\"u}rr and Teufel, 2009]{Durr:2009fk}
D{\"u}rr, D. and Teufel, S. (2009).
\newblock {\em Bohmian mechanics: the physics and mathematics of quantum
  theory}.
\newblock Berlin: Springer.

\bibitem[Egg and Esfeld, 2014]{Egg:2014aa}
Egg, M. and Esfeld, M. (2014).
\newblock Non-local common cause explanations for {EPR}.
\newblock {\em European Journal for Philosophy of Science}, 4:181--196.

\bibitem[Egg and Esfeld, 2015]{Egg:2015aa}
Egg, M. and Esfeld, M. (2015).
\newblock Primitive ontology and quantum state in the {GRW} matter density
  theory.
\newblock {\em Synthese}, 192(10):3229--3245.

\bibitem[Einstein, 1948]{Einstein:1948aa}
Einstein, A. (1948).
\newblock Quanten-{M}echanik und {W}irklichkeit.
\newblock {\em Dialectica}, 2:320--324.

\bibitem[Einstein et~al., 1935]{Einstein:1935aa}
Einstein, A., Podolsky, B., and Rosen, N. (1935).
\newblock Can quantum-mechanical description of physical reality be considered
  complete?
\newblock {\em Philosophical Review}, 47:777--780.

\bibitem[Esfeld, 2014a]{Esfeld:2014ac}
Esfeld, M. (2014a).
\newblock The primitive ontology of quantum physics: guidelines for an
  assessment of the proposals.
\newblock {\em Studies in History and Philosophy of Modern Physics},
  47:99--106.

\bibitem[Esfeld, 2014b]{Esfeld:2014aa}
Esfeld, M. (2014b).
\newblock Quantum {H}umeanism, or: physicalism without properties.
\newblock {\em The Philosophical Quarterly}, 64(256):453--470.

\bibitem[Esfeld and Gisin, 2014]{Esfeld:2014ad}
Esfeld, M. and Gisin, N. (2014).
\newblock The {GRW} flash theory: a relativistic quantum ontology of matter in
  space-time?
\newblock {\em Philosophy of Science}, 81:248--264.

\bibitem[Esfeld et~al., 2014]{Esfeld:2014ab}
Esfeld, M., Lazarovici, D., Hubert, M., and D{\"u}rr, D. (2014).
\newblock The ontology of {B}ohmian mechanics.
\newblock {\em British Journal for the Philosophy of Science}, 65(4):773--796.

\bibitem[Feynman et~al., 1963]{Feynman:1963aa}
Feynman, R.~P., Leighton, R.~B., and Sands, M. (1963).
\newblock {\em The {F}eynman lectures on physics. {V}olume 1}.
\newblock Reading (Massachusetts): Addison-Wesley.

\bibitem[Forrest, 1988]{Forrest:1988aa}
Forrest, P. (1988).
\newblock {\em Quantum metaphysics}.
\newblock Oxford: Blackwell.

\bibitem[Ghirardi et~al., 1995]{Ghirardi:1995aa}
Ghirardi, G.~C., Grassi, R., and Benatti, F. (1995).
\newblock Describing the macroscopic world: closing the circle within the
  dynamical reduction program.
\newblock {\em Foundations of Physics}, 25(1):5--38.

\bibitem[Ghirardi et~al., 1990]{Ghirardi:1990aa}
Ghirardi, G.~C., Pearle, P., and Rimini, A. (1990).
\newblock Markov processes in {H}ilbert space and continuous spontaneous
  localization of systems of identical particles.
\newblock {\em Physical Review A}, 42:78--89.

\bibitem[Ghirardi et~al., 1986]{Ghirardi:1986aa}
Ghirardi, G.~C., Rimini, A., and Weber, T. (1986).
\newblock Unified dynamics for microscopic and macroscopic systems.
\newblock {\em Physical Review D}, 34(2):470--491.

\bibitem[Lewis, 1997]{Lewis:1997aa}
Lewis, P.~J. (1997).
\newblock Quantum mechanics, orthogonality, and counting.
\newblock {\em British Journal for the Philosophy of Science}, 48:313--328.

\bibitem[Maudlin, 1995]{Maudlin:1995aa}
Maudlin, T. (1995).
\newblock Three measurement problems.
\newblock {\em Topoi}, 14:7--15.

\bibitem[Maudlin, 2010]{Maudlin:2010aa}
Maudlin, T. (2010).
\newblock Can the world be only wave-function?
\newblock In Saunders, S., Barrett, J., Kent, A., and Wallace, D., editors,
  {\em Many worlds? Everett, quantum theory, and reality}, pages 121--143.
  Oxford: Oxford University Press.

\bibitem[Maudlin, 2011]{Maudlin:2011aa}
Maudlin, T. (2011).
\newblock {\em Quantum non-locality and relativity. Third edition}.
\newblock Chichester: Wiley-Blackwell.

\bibitem[Miller, 2014]{Miller:2014aa}
Miller, E. (2014).
\newblock Quantum entanglement, {B}ohmian mechanics, and {H}umean
  supervenience.
\newblock {\em Australasian Journal of Philosophy}, 92:567--583.

\bibitem[Monton, 2004]{Monton:2004aa}
Monton, B. (2004).
\newblock The problem of ontology for spontaneous collapse theories.
\newblock {\em Studies in History and Philosophy of Modern Physics},
  35(3):407--421.

\bibitem[Mundy, 1989]{Mundy:1989aa}
Mundy, B. (1989).
\newblock Distant action in classical electromagnetic theory.
\newblock {\em British Journal for the Philosophy of Science}, 40(1):39--68.

\bibitem[Norsen, 2005]{Norsen:2005aa}
Norsen, T. (2005).
\newblock Einstein's boxes.
\newblock {\em American Journal of Physics}, 73:164--176.

\bibitem[Norsen, 2014]{Norsen:2014aa}
Norsen, T. (2014).
\newblock The pilot-wave perspective on spin.
\newblock {\em American Journal of Physics}, 82(4):337--348.

\bibitem[Norsen et~al., 2015]{Norsen:2015aa}
Norsen, T., Marian, D., and Oriols, X. (2015).
\newblock Can the wave function in configuration space be replaced by
  single-particle wave functions in physical space?
\newblock {\em Synthese}, 192:3125--3151.

\bibitem[Oldofredi et~al., 2016]{Oldofredi:2016aa}
Oldofredi, A., Lazarovici, D., Deckert, D.-A., and Esfeld, M. (2016).
\newblock From the universe to subsystems: Why quantum mechanics appears more
  stochastic than classical mechanics.
\newblock {\em Fluctuations and Noise Letters}, 15(3):164002: 1--16.

\bibitem[Pylkk{\"a}nen et~al., 2015]{Pylkkanen:2014aa}
Pylkk{\"a}nen, P., Hiley, B.~J., and P{\"a}ttiniemi, I. (2015).
\newblock Bohm's approach and individuality.
\newblock In Guay, A. and Pradeu, T., editors, {\em Individuals across the
  sciences}, chapter~12, pages 226--246. Oxford: Oxford University Press.

\bibitem[Schr{\"o}dinger, 1935]{Schroedinger:1935aa}
Schr{\"o}dinger, E. (1935).
\newblock Die gegenw{\"a}rtige {S}ituation in der {Q}uantenmechanik.
\newblock {\em Naturwissenschaften}, 23:807--812.

\bibitem[Su{\'a}rez, 2015]{Suarez:2015aa}
Su{\'a}rez, M. (2015).
\newblock Bohmian dispositions.
\newblock {\em Synthese}, 192:3203--3228.

\bibitem[Tumulka, 2006]{Tumulka:2006aa}
Tumulka, R. (2006).
\newblock A relativistic version of the {G}hirardi--{R}imini--{W}eber model.
\newblock {\em Journal of Statistical Physics}, 125(4):821--840.

\bibitem[Tumulka, 2011]{Tumulka:2011aa}
Tumulka, R. (2011).
\newblock Paradoxes and primitive ontology in collapse theories of quantum
  mechanics.
\newblock {\em http://arxiv.org/quant-ph/1102.5767v1}.

\bibitem[Vassallo et~al., 2016]{Vassallo:2016ac}
Vassallo, A., Deckert, D.-A., and Esfeld, M. (2016).
\newblock Relationalism about mechanics based on a minimalist ontology of
  matter.
\newblock {\em European Journal for Philosophy of Science}, pages DOI
  10.1007/s13194--016--0160--2, preprint
  http://philsci--archive.pitt.edu/12398/, arXiv:1609.00277[physics.hist--ph].

\bibitem[Vassallo and Ip, 2016]{Vassallo:2016aa}
Vassallo, A. and Ip, P.~H. (2016).
\newblock On the conceptual issues surrounding the notion of relational
  {B}ohmian dynamics.
\newblock {\em Foundations of Physics}, 46(8):943--972.

\bibitem[Wallace, 2008]{Wallace:2008aa}
Wallace, D. (2008).
\newblock Philosophy of quantum mechanics.
\newblock In Rickles, D., editor, {\em The {A}shgate companion to contemporary
  philosophy of physics}, pages 16--98. Aldershot: Ashgate.

\bibitem[Wallace, 2014]{Wallace:2014aa}
Wallace, D. (2014).
\newblock Life and death in the tails of the {GRW} wave function.
\newblock {\em arXiv:1407.4746 [quant-ph]}.

\end{thebibliography}

\end{document}